\documentstyle[psfig]{mn}

\title[The triple degenerate star WD\,1704+481]
{The triple degenerate star WD\,1704+481}
\author[P. F. L. Maxted et~al.]
       {P. F. L. Maxted$^1$, T. R. Marsh$^1$, C.~K.~J.~Moran$^1$ and
Z.~Han$^{2,3,4}$\\
$^1$    University of Southampton, Department of Physics \& Astronomy,
        Highfield, Southampton, S017 1BJ, UK \\
$^2$    Yunnan Observatory, Academia Sinica, Kunming, 650011, P.R. China\\
$^3$    National Astronomical Observatories, the Chinese Academy of Sciences\\
$^4$    Institute of Astronomy, Madingley Road, Cambridge, CB3 0HA, UK
}
\date{Accepted 1999 
      Received 1999 }

\pagerange{\pageref{firstpage}--\pageref{lastpage}}
\pubyear{1999}

\newcommand{\Msolar}{\mbox{${\rm M}_{\odot}\,$}}
\newcommand{\Rsolar}{\mbox{${\rm R}_{\odot}\,$}}
\newcommand{\kms}{\mbox{${\rm km\,s}^{-1}$}}
\begin{document}

\maketitle

\label{firstpage}

\begin{abstract}
 WD\,1704+481 is a visual binary in which both components are white dwarfs. We
present spectra of the H$\alpha$ line of both stars which show that one
component (WD\,1704+481.2 = Sanduleak~B = GR\,577) is a close binary
with two white dwarf components. Thus, WD\,1704+481 is the first known
triple degenerate star. From radial velocity measurements of the close
binary we find an orbital period of 0.1448d, a mass ratio, $q=\frac{
M_{\rm bright}}{M_{\rm faint}}$ of
$q=0.70\pm$0.03 and a difference in the gravitational redshifts of
11.5$\pm$2.3\kms. The masses of the close pair of white dwarfs predicted
by the mass ratio and gravitational redshift difference combined with
theoretical cooling curves are 0.39$\pm$0.05\Msolar\ and
0.56$\pm$0.07\Msolar. WD\,1704+481 is therefore also likely to be the
first example of a double degenerate in which the less massive white dwarf is
composed of helium and the other white dwarf is composed of carbon and oxygen. 
\end{abstract} 

\section{Introduction}
 WD\,1704+481 was identified as a pair of white dwarfs of
similar brightness separated by about 6\,arcscec by Sanduleak and Pesch
(1982) from objective prism plates. Spectrophotometry of the pair by
Greenstein, Dolez \& Vauclair (1983) showed that, although their visual
magnitudes are almost identical, the SE component (WD\,1704+481.1 =
Sanduleak~A = GR\,576, V=14.48, (B$-$V)=-0.09) is bluer and, therefore, hotter
than the NW component (WD\,1704+481.2 = Sanduleak~B = GR\,577, V=14.45,
(B$-$V)=0.14). This immediately suggests that the cooler component must have a
larger radius which also implies a lower mass for a cool white dwarf such as
Sanduleak~B. Greenstein et~al. estimated a mass of 0.32\,--\,0.43\Msolar\  for
Sanduleak~B which is well below the typical mass of white dwarfs (0.55\Msolar,
Bergeron, Saffer \& Liebert 1992).

 Low mass white dwarfs such as Sanduleak~B are thought to be the result of
binary star evolution, in which the evolution of a star during the red giant
phase is interrupted by interactions with a nearby  star. The physics of this
interaction is complex but it is thought to lead to the stripping of the outer
hydrogen layers from the red giant in a ``common-envelope'' phase, halting the
formation of the degenerate helium core and leading to the formation of an
anomalously low mass white dwarf (Iben \& Livio 1993). The hypothesis that
binary star evolution forms low mass white dwarfs was confirmed by the
discovery by Marsh, Dhillon \& Duck (1995) of at least 5 short period binary
white dwarfs in a sample of 7 low mass white dwarfs.  

 In this paper we present  spectra of the H$\alpha$ line of Sanduleak~B which
clearly show that it is a close binary with two white dwarf components -- a
double degenerate star (DD). We derive the spectroscopic orbits of both
components and show that it is likely to be the first known example of a binary
white dwarf with one white dwarf composed of helium and one composed of carbon
and oxygen.

\section{Observations and reductions.}
 The data for this study come from observations obtained with the 2.5m Isaac
Newton Telescope (INT) in September 1998 and the 4.2m William  Herschel
Telescope (WHT) in October 1998.  The INT spectra were obtained with the
intermediate dispersion spectrograph using the 500mm camera, a 1200 line/mm
grating, a 0.9arcsec slit  and a TEK charge coupled device (CCD) as a
detector at a dispersion of 0.39\AA\ per pixel. The WHT spectra
were obtained using the blue arm of the ISIS spectrograph, a 1200 line/mm
grating, a 0.83arcsec slit and an EEV CCD at a dispersion of 0.22\AA\ per
pixel. Integration times varied between 600s and 1200s. Each observation of
the stars was bracketed by observations of a copper-neon arc. We set
the angle of the slit so that spectra of both stars were obtained
simultaneously. The position angle estimated from this slit angle
was 295$^{\circ}$ for the INT and 290$^{\circ}$ for the WHT, which
agrees well with the  PA of 289$^{\circ}\pm 5^{\circ}$ given by
Greenstein et~al.  The separation of the stars measured from our
INT images is 6.0\,arcsec with an uncertainty of a few tenths of an
arcsecond. The seeing was good for all the observations ($\sim$ 1arcsec)
so the spectra of the stars are clearly separated in all our images. Of
the 49 spectra, 41 were obtained with the INT.

 The bias level in all the images determined from the clipped-mean in the
overscan region was subtracted from all the images before further processing.
For the INT spectra, several images of a tungsten lamp were combined
to form a normalized master flat-field image for each night's data. Similar
images for the WHT spectra show mild fringing. We therefore combined flat-field
images taken immediately before and after each observation of the stars in
order to form a normalized flat-field image for each image of the star.
Extraction of the spectra from the images was performed automatically using
optimal extraction to maximize the signal-to-noise of the resulting spectra
(Horne 1986). Uncertainties due to photon statistics are rigorously
followed through the data reduction process so that reliable uncertainties are
known for every point in the final spectra. The arcs associated with each
stellar spectrum were extracted using the same weighting determined for the
stellar image to avoid possible systematic errors due to tilted spectra.
The wavelength scale was determined from a fourth-order polynomial fit
to measured arc-line positions. We calibrated the wavelength response of
the WHT spectra using observations of the standard star
BD\,+33$^{\circ}$~2642 (Bohlin 1996). To calibrate the wavelength
response  of the INT spectra we used a least-squares fit of a smooth
function to the ratio of the average WHT spectrum of WD\,1740+481.2 and
the  average INT spectrum of the same star. The core of the H$\alpha$
line and  regions affected by telluric absorption were excluded from the
fit.  We then normalized all the spectra using a linear fit to the
continuum either side of the H$\alpha$ line. 

We measured the resolution of the spectra by fitting a Gaussian profile by
least-squares to a neon arc line at 6532\AA\ in a series of spectra.  The
full-width at half-maximum (FWHM) of the model profile was typically 0.94\AA\
for the INT spectra and 0.45\AA\ for the WHT spectra. The INT spectra for one
night are affected by poor spectrograph focus and have a resolution of
1.2\AA.

\section{Measurement of the radial velocities.}
\begin{figure*}
\leavevmode\centering{
\psfig{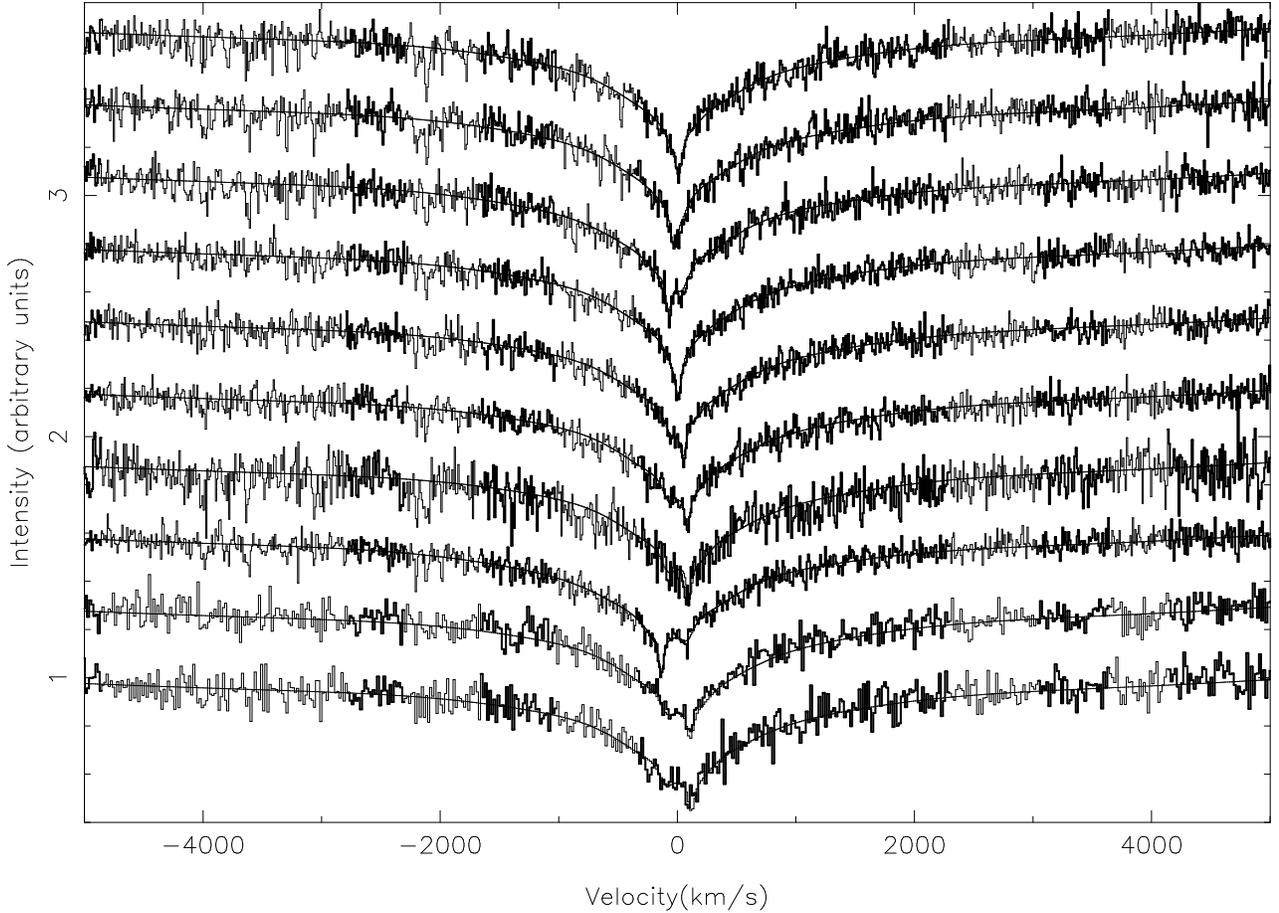}}
\caption{\label{SpecFig} Examples of our WD\,1704+481.2 spectra and the
least-squares fit used to derive the orbital parameters of the binary.
The spectra (which are in no particular order) are offset  vertically from
each other for clarity and the fit is shown as a smooth line. The regions of
the spectrum affected by telluric absorption and excluded from the fit are
plotted with faint lines.
}
\end{figure*}

 Our spectra of the cooler component of the visual binary  clearly show
two sharp cores to the H$\alpha$ line (Fig.~\ref{SpecFig}) which vary in
position periodically in a sinusoidal manner, indicating the presence of
two white dwarfs in a close binary orbit. An initial estimate of the
period was obtained by measuring the radial velocity of the deeper core
using a model profile. We first used a simultaneous least-squares fit
to all the spectra of four Gaussian profiles  with a common position but
independent depths and widths to determine the model profile. The shape
of the profile was then fixed in a least-squares fit to the individual
spectra in which  only the position of the model profile was allowed to
vary. These initial velocities are affected by the presence of the weaker
core, but were good enough to determine the orbital period. A
periodogram of these radial velocities showed a clear peak at 6.91
cycles per day, which confirmed our initial impression that the period
is around 0.145d. 

 To measure the radial velocities of the two components more precisely we
used a simultaneous fit to all the spectra of two model profiles, one for
each star, in which the position of each model profile is predicted from a
circular orbit of the form $\gamma + K\sin(2\pi(T-T_0)/P)$. In this way we are
able to determine the shape of the two profiles and the parameters of
the two circular orbits directly. The (variable) resolution of each of
the spectra and the effects of smearing due to orbital motion are
included in the fitting process. There are many free-parameters in this
fitting process so we used a series of least-squares fits in which first
the profile shapes were fixed while the parameters of the orbit were
varied and then {\it vice versa}, until we had established values for
all the parameters which were nearly optimal. Only data within
5000\,km\,s$^{-1}$ of H$\alpha$ and unaffected by telluric absorption are
included in the fitting process. We used four Gaussian profiles to model
the broad wings of the H$\alpha$ line and the core of star with the
deeper core and two Gaussian profiles for the other star. A  polynomial
is also included in the fitting process to allow for smooth, asymmetric
features in the profile. For the final least-squares fit the parameters
of the profile shapes and the orbit were all varied independently. The
parameters of this final fit are given in Table~\ref{FitTable}. We
designate the star with the deeper H$\alpha$ core component B and the
star with the weaker H$\alpha$ core is then component C as we will refer
to Sanduleak~A as component A throughout to avoid confusion.

 We also measured the radial velocities of Sanduleak~A using the same method
employed to measure the initial radial velocities of Sanduleak~B. There is no
significant variability in these radial velocities, which have an
average value of $\gamma_{\rm A} = 0.6 \pm 0.3\kms$.

\section{Discussion}

\begin{figure}
\leavevmode\centering{
\psfig{file=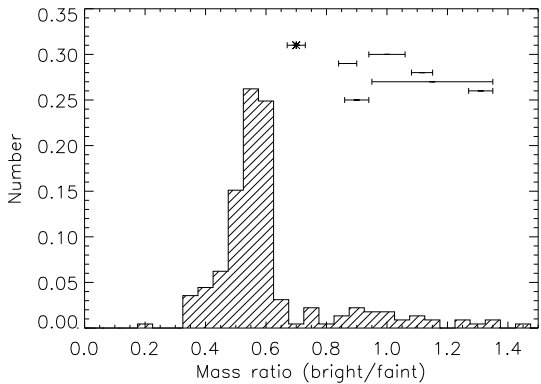,width=0.45\textwidth} }
\caption{\label{QFig} The predicted mass ratio distribution for DDs with
measured mass ratios from the model of Han (1998). Measured mass ratios
for double degenerates, including WD\,1704+481.2 (marked with an
asterisk) are indicated by error bars.}
\end{figure}

 The first obvious result of our analysis is the mass ratio $q=\frac{M_{\rm
B}}{M_{\rm C}} = 0.70 \pm 0.03$. WD\,1704+481.2 is one of several double
degenerates (DDs) which have been identified among low mass white dwarfs
($M\la0.49$\Msolar) where the fainter companion is sufficiently young, i.e,
hot, for its H$\alpha$ core to be detected and a mass ratio derived (Moran,
Marsh \& Maxted, 1999). These
observed mass ratios for DDs are shown in Fig.~\ref{QFig}. Component C has the
weaker core in our H$\alpha$ spectra. The depth of the core does not vary
strongly with temperature over the temperature range expected for components B
and C (Koester et~al. 1998), so we can confidently state that component C is
the fainter component.  For comparison, we also plot the results of Han
(1998). We applied a set of selection criteria to the simulated population of
white dwarf binaries to approximate these selection effects, namely, that the
younger white dwarf is less massive than 0.49\Msolar and that the older white
dwarf is younger than 1\,Gyr.  The mass ratio distribution for white dwarf
binaries predicted by the models of Iben, Tutukov \& Yungelson (1997) using
similar selection criteria but their own ``standard model'' is quite similar,
i.e. it shows a single peak near $q=0.7$. WD\,1704+481.2 lies near the peak of
the theoretical distribution, which is a success for the theoretical model.
What is less clear is why the majority of DDs with measured mass ratios do not
have mass ratios which accord with the models.

\begin{figure}
\leavevmode\centering{
\psfig{file=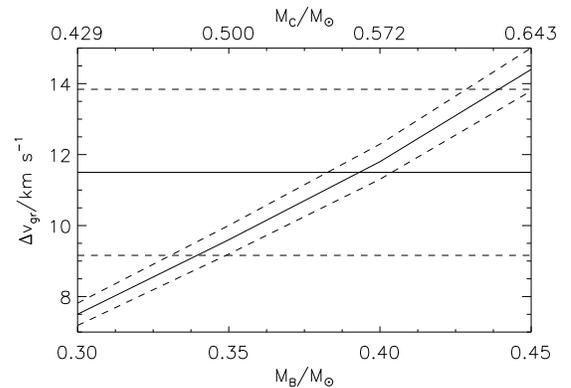,width=0.45\textwidth} }
\caption{\label{GFig} The predicted gravitational redshift difference, $\Delta
v_{\rm gr}$, as a function of
the component masses (solid diagonal line) and its uncertainty due to the 
uncertainty in the mass ratio (dashed diagonal lines). The observed value of
 $\Delta v_{\rm gr}$ (solid horizontal line)  and its uncertainty (horizontal
dashed line) are also shown.}
\end{figure}
\begin{figure}
\leavevmode\centering{
\psfig{file=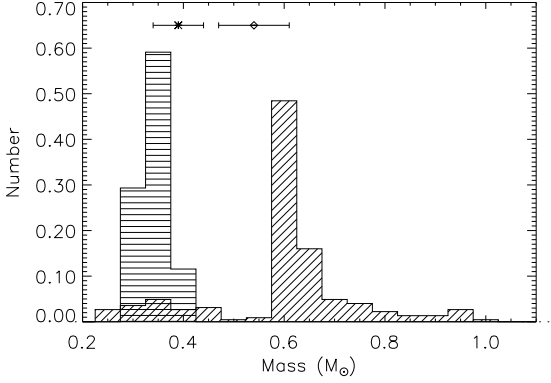,width=0.45\textwidth} }
\caption{\label{MFig} The predicted mass distributions  for the brighter
component (horizontal hatching) and fainter component (diagonal
hatching) of DDs with measured mass ratios from the model of Han (1998).
The observed masses for the brighter component (asterisk) and fainter
component (diamond) of  WD\,1704+481.2 are also shown. Note that the
uncertainties shown are correlated, i.e., an increased mass for the brighter
component implies an increased mass for the fainter component.} \end{figure}

 We can use the observed difference $\gamma_{\rm B} - \gamma_{\rm C} =
-11.5\pm2.3 \kms$ to determine the masses of the stars as follows. The
gravitational redshift of the light from the stars is given by $v_{\rm gr} =
0.635 (M/\Msolar)/(R/\Rsolar) \kms$. Since one star is more massive than the
other and more massive white dwarfs have smaller radii (for a given
temperature), the more massive component will appear to have a larger (more
positive) systemic velocity, as is observed. To determine the radius of
component B we use the models of Althaus \& Benvenuto (1997) for helium
white dwarfs. We assume a temperature of 9500K for component B
(Greenstein, Dolez \& Vauclair 1983) and use parabolic interpolation to find
the radius. Note that a 1000K difference in temperature changes the radius
derived by less than one percent so the temperature assumed has a negligible
effect on our results. We assume that component C is slightly cooler (8500K)
and use the mixed-composition models of Wood (1990) to determine its radius.
In Fig.~\ref{GFig} we show the predicted gravitational redshift difference,
$\Delta v_{\rm gr}$ as a function of the mass of each component. The observed
value of $\Delta v_{\rm gr}$ is also indicated and we see that this observed
value implies $M_{\rm B}=0.39 \pm 0.05\Msolar$ and $M_{\rm
C}=0.56\pm 0.07\Msolar$. White dwarfs more massive than about
0.5\Msolar will have  passed through a helium-burning stage and have a
carbon-oxygen composition, so these masses suggest that component B is a white
dwarf composed of helium and component  C is composed of carbon and oxygen. In
Fig.~\ref{MFig} we see that the masses of the two components are roughly what
would be expected from the theoretical models of Han (1998).

 The mass of component C is typical for single white dwarfs (Bergeron, Saffer
\& Liebert 1992) and component A has the same gravitational redshift.  The
change in radius with temperature in this regime is very small, so we expect
component A has a normal mass. Component A is $\ga$200\,AU distant from
components B and C (Greenstein et~al. 1983) and the initial separation of
components B and C must have been smaller than $\sim 1$\,AU for a
common-envelope phase to have occurred.  The ratio of the orbital periods is
sufficiently large that we need not worry about dynamical instability of the
orbits (Kiseleva, Eggleton \& Anosova 1994) and so we can be confident that
the evolution of the inner binary will have been unaffected by the presence of
component A and {\it vice versa}.

 From the parameters of the orbit we find $M_{\rm B}\sin^3i  =
0.075\pm0.003\Msolar$ and $M_{\rm C}\sin^3i  = 0.108\pm0.006\Msolar$
which, combined with the masses derived above, yields an inclination of
$61^{\circ}$. The separation of components B and C is only 0.74\Rsolar\ 
but the small radii of white dwarfs ($\sim0.01\Rsolar$) rules out the
possibility of observable eclipses in this binary.

\section{Conclusion}
 
 We have shown that WD\,1704+481 is a hierarchical triple star in which all
three components are white dwarfs, i.e., a triple degenerate star. The
outermost star, component A appears to be a typical white dwarf. Components B
and C are a close binary with an orbital period of only 0.145d. The mass ratio
($q=\frac{M_{\rm bright}}{M_{\rm faint}}=\frac{M_{\rm B}}{M_{\rm
C}}=0.70\pm0.03$) for the inner binary is the first measured mass ratio for a
double degenerate which is close to the peak of the mass ratio distribution
predicted by theoretical models. Similarly, the  masses of components B and C
derived from the difference between their gravitational redshifts ($M_{\rm
B}=0.39\pm 0.05\Msolar$ and  $M_{\rm C}=0.56\pm 0.07\Msolar$)
appear near the peaks of the theoretical mass distributions. These suggest
that the component C is a typical white dwarf composed of carbon and oxygen
and that component B is composed of helium. 

\begin{table}
\caption{\label{FitTable} Orbital parameters of the final least-squares fit to
 the spectra of WD\,1704+481.2. The full-width at half-minimum (FWHM) and
depths (D) of each of the Gaussian profiles used to model the H$\alpha$
line are also given. Other symbols are defined in the text.}
\begin{tabular}{lr}
 Circular orbit \\
 HJD$(T_0)-2451000$    &   77.4070     $\pm$   0.0002    \\
 $P (d)$                 &   0.1447864   $\pm$   0.0000025 \\
 $\gamma_{\rm B}(\kms)$ &  $-$13.0        $\pm$   0.8       \\
 $K_{\rm B}(\kms)$      &  $-$135.4       $\pm$   1.3       \\
 $\gamma_{\rm C}(\kms)$ &  $-$1.4         $\pm$   2.2       \\
 $K_{\rm C}(\kms)$      &   94.5        $\pm$   3.8       \\
 Component B           \\
  FWHM$_1$ (\AA) &  0.86 $\pm$   0.09 \\
  D$_1$          & 0.147 $\pm$   0.009 \\
  FWHM$_2$ (\AA) &  5.8 $\pm$   0.4 \\
  D$_2$          & 0.078 $\pm$   0.004 \\
  FWHM$_3$ (\AA) &  21.6$\pm$   1.1 \\
  D$_3$          & 0.116 $\pm$   0.007 \\
  FWHM$_4$ (\AA) &  55.0$\pm$   1.7 \\
  D$_4$          & 0.138 $\pm$   0.005 \\
 Component C \\
  FWHM$_1$ (\AA) &   2.5 $\pm$   0.2 \\
  D$_1$        & 0.063 $\pm$   0.003 \\
  FWHM$_2$ (\AA) &  16.7 $\pm$   1.1 \\
  D$_2$        & 0.055 $\pm$   0.004 \\
\end{tabular}
\end{table}

\section*{Acknowledgements}
 PFLM was supported by a PPARC post-doctoral grant. CM was supported by a
PPARC post-graduate studentship. The William Herschel Telescope and the Isaac
Newton Telescope are operated on the island of La Palma by the Isaac Newton
Group in the Spanish Observatorio del Roque de los Muchachos of the Instituto
de Astrofisica de Canarias.

\label{lastpage}

\end{document}